\documentclass{eptcs}
\providecommand{\event}{F-IDE 2016, 3rd Workshop on Formal Integrated Development Environment\\
Satellite event of the 21st International Symposium on Formal Methods (FM2016)} % Name of the event you are submitting to
\usepackage{breakurl}             % Not needed if you use pdflatex only.
\usepackage{underscore}           % Only needed if you use pdflatex.

\usepackage[english]{babel}
\usepackage[utf8]{inputenc}
\usepackage[T1]{fontenc}
\usepackage{amssymb}
\usepackage{graphicx}
\usepackage{subfig}
\usepackage{booktabs}
\usepackage{color}
\definecolor{dkgreen}{rgb}{0,0.6,0}
\definecolor{gray}{rgb}{0.5,0.5,0.5}
\definecolor{mauve}{rgb}{0.58,0,0.82}
\usepackage{pdflscape} %Gives us landscape pages with begin{landscape}. Use this for pages with very large diagrams.
\usepackage{mwe}

\usepackage{pifont}
\usepackage{alltt}
\usepackage{amsmath,amsfonts}

\newcounter{MALcounter}

\newenvironment{Mbit}
{\begin{tabbing}x \= x \= $\text{[vis]}\ $\=xxxxxxxx \= xx\kill}
	{\end{tabbing}}

\newcommand{\Mword}[1]{\ensuremath{\mathit{#1}}}

\newcommand{\Mplain}[1]{\>\>$\mathit{#1}$}

\usepackage{oz}
\usepackage{tcolorbox}
\usepackage{mathtools}

\usepackage{hyperref}
\urldef{\mailsa}\path|emails|

\usepackage{tabularx}

\title{Evaluation of Formal IDEs for Human-Machine Interface Design and Analysis: The Case of CIRCUS and PVSio-web}%\vspace*{-12pt}}

\author{
	Camille Fayollas$^{1}$ \quad C\'elia Martinie$^{1}$ \quad Philippe Palanque$^{1}$ \quad Paolo Masci$^{2}$\vspace*{-0.5pt}
	\and \quad Michael D. Harrison$^{2,3}$ \quad Jos\'e C. Campos$^{2}$ \quad Saulo Rodrigues e Silva$^{2}$
	\smallskip
	\institute{$^{1}$ICS-IRIT, University of Toulouse, Toulouse, France}\vspace*{-2pt}
	\email{\{fayollas,martinie,palanque\}@irit.fr}
	\smallskip
	\institute{$^{2}$HASLab/INESC TEC and Universidade do Minho, Braga, Portugal}\vspace*{-2pt}
	\email{\{paolo.masci,jose.c.campos,saulo.r.silva\}@inesctec.pt}
	\smallskip
	\institute{$^{3}$Newcastle University, Newcastle upon Tyne, United Kingdom}\vspace*{-2pt}
	\email{michael.harrison@newcastle.ac.uk}
}

\def\titlerunning{Evaluation of Formal IDEs for Human-Machine Interface Design and Analysis}
\def\authorrunning{C. Fayollas et al.}

\begin{document}
\maketitle

\begin{abstract}
Critical human-machine interfaces are present in many systems including avionics systems and medical devices. Use error is a concern in these systems both in terms of hardware panels and input devices, and the software that drives the interfaces. Guaranteeing safe usability, in terms of  buttons, knobs and displays is now a key element in the overall safety of the system. New integrated development environments (IDEs) based on formal methods technologies have been developed by the research community to support the design and analysis of high-confidence human-machine interfaces. To date, little work has focused on the comparison of these particular types of formal IDEs. This paper compares and evaluates two state-of-the-art toolkits: CIRCUS, a model-based development and analysis tool based on Petri net extensions, and PVSio-web, a prototyping toolkit based on the PVS theorem proving system.
\end{abstract}

\section{Introduction}
Use error is a major concern in critical interactive systems. Consider, for example, a pilot calibrating flight instruments before take-off. When calibrating the barometer used to measure the aircraft's altitude, a consistency check should be performed automatically by the cockpit software to help guard against use errors, such as mistyping a value or selecting the wrong units. 

New IDEs based on formal methods have been developed by the research community to support the design and analysis of high-confidence human-machine interfaces. Each IDE supports different types of analysis, ranging from functional correctness (e.g., absence of deadlocks and coding errors such as division by zero) to compliance with usability and safety requirements (e.g., assessing the response to user tasks, or the visibility of critical device modes). Choosing the right tool is important to ensure efficiency of the analysis and that the analysis addresses the appropriate safety concerns relating to use. To date, little work has been done to compare and evaluate different formal IDEs for human-machine interface design and analysis, and little or no guidance is available for developers to understand which IDE can be used most effectively for which kind of analysis. This paper describes a first step towards addressing this gap.

\smallskip\noindent
{\bf Contribution.} We compare and evaluate two state-of-the-art formal verification technologies for the analysis of human-machine interfaces: CIRCUS \cite{edcc2014}, a model-based development and analysis tool that uses Petri net extensions; and PVSio-web \cite{cav2015}, a prototyping toolkit based on the PVS theorem prover. The aim of this work is to provide guidance to developers to understand which tool can be used most effectively for which kind of analysis of interactive systems. Both tools have their foundations in existing formal technologies, but are focused towards particular issues relating to the user interface. The capabilities of the two tools are demonstrated in the paper through a common case study based on a critical subsystem in the cockpit of a large civil aircraft. A taxonomy is developed as a result of the comparison that can be used to describe the characteristics of other similar tools.

\smallskip\noindent
{\bf Organisation.} The remainder of the paper is organised as follows. Section \ref{sec:fide} illustrates typical features of formal IDEs for the design and analysis of human-machine interfaces, and presents a detailed description of CIRCUS and PVSio-web. Section~\ref{sec:example} introduces the common example for comparison of the selected tools, as well as the developed models. Section~\ref{sec:comparison} presents the metrics for comparing the IDEs, and then uses the metrics as a basis for the comparison. Section~\ref{sec:conclu} concludes the paper and presents future directions in which the tools may evolve.

\section{The formal modelling and analysis of user interfaces}\label{sec:fide}

Formal tools for the modelling and analysis of human-machine interfaces are designed to support multi-disciplinary teams of designers from different engineering disciplines, including human factors engineering (to establish usability requirements, run user studies and interpret compliance), formal methods (to verify compliance of a system design with design requirements), and software engineering (to develop prototypes and software code, e.g., using model-based development methods). Although several tools provide graphical model editors and automated functions for modelling and analysis of interactive elements of a system, different tools are usually complementary, as they support different levels of description, and different types of analysis, ranging from micro-level aspects of human-machine interaction, e.g., aspect and behaviour of user interface widgets, to the analysis of the wider socio-technical system within which the interactive system is used.

In the present work, we compare two state-of-the-art formal tools developed by two different research teams: CIRCUS~\cite{edcc2014}, a toolkit for model-based development of interactive systems; and PVSio-web~\cite{cav2015}, a toolkit for model-based development of user interface software. Both build on tools that have been developed more generally for model based design and software engineering, extending them with features that are particularly useful when considering the human-machine interface or the wider socio-technical system.

Other formal tools that could be used (and in some cases have been used) for the analysis of human-machine interfaces exist. They offer functionalities that complement those of CIRCUS and PVSio-web. The evaluation of these other tools is not within the scope of this paper, although
a brief overview can be found in Section~\ref{sec:others}.

\subsection{CIRCUS}

CIRCUS, which stands for Computer-aided-design of Interactive, Resilient, Critical and Usable Systems, is an IDE for the formal verification of the system's behaviour as well as the analysis of compatibility between the user's task and the system's behaviour. CIRCUS includes three tools:

\begin{itemize}
	\item {\bf HAMSTERS} (Human-centred Assessment and Modelling to Support Task Engineering for Resilient Systems) is a tool for editing and simulating task models. The tool can be used to ensure consistency, coherence, and conformity between assumed or prescribed user tasks and the sequence of actions necessary to operate interactive systems \cite{barboni2010beyond}. The notation used in the tool makes it possible to structure users' goals and sub-goals into hierarchical task trees. Qualitative temporal relationships among tasks are described by operators. Various notational elements support modelling of specialised task types, explicit representations of data and knowledge, device descriptions, genotypes and phenotypes of errors, and collaborative tasks.
	\item {\bf PetShop} (Petri Net workshop) is a tool for creating, editing, simulating and analysing system models using the ICO (Interactive Cooperative Objects) notation \cite{navarre2009icos,martinie2011structuring}. The ICO notation allows developers to specify the behaviour of interactive systems. The notation uses Petri Nets for describing dynamic behaviours, and uses object-oriented concepts (including dynamic instantiation, classification, encapsulation, inheritance and client/server relationships) to describe structural or static aspects of the system.
	\item {\bf SWAN} (Synergistic Workshop for Articulating Notations) is a tool for the co-execution of PetShop models and HAMSTERS models \cite{barboni2010beyond}. The tool allows developers to establish correspondences between system behaviours and tasks, and perform automated system testing by means of co-execution~\cite{eics2016}.
\end{itemize}

\subsection{PVSio-web}
PVSio-web is a toolkit for model-based development of user interface software. The toolkit is based on and extends an established theorem prover, PVS \cite{pvs}, providing a graphical environment for constructing, visualising and analysing formal models of user interface software. PVSio-web has three main components:
\begin{itemize}
	\item {\bf Prototype Builder and Simulator}. This tool allows developers to create device prototypes based on formal models, and run them within a Web browser. The visual aspect of the prototype uses an interactive picture of the device. Developers create programmable areas over the picture to identify input widgets (e.g., buttons) and output widgets (e.g., displays, LEDs). The tool automatically translates user actions over input widgets (e.g., button presses) into PVS expressions that can be evaluated within PVSio~\cite{munoz2003rapid}, the native PVS component for animating executable PVS models. The Simulator tool executes PVSio in the background, and the effects of the execution are automatically rendered using the output widgets of the prototype to closely resemble the visual appearance of the real system in the corresponding states.
	
	\item {\bf Emucharts Editor}. This tool facilitates the creation of formal models using visual diagrams known as Emucharts. These diagrams are based on Statecharts \cite{horrocks1999constructing}. The tool allows developers to define the following design elements: states, representing the different modes of the system; state variables, representing the characteristics of the system state; and transitions, representing events that change the system state. The tool incorporates a model generator that translates the Emucharts diagram into executable PVS models automatically. The model generator also supports the generation of other different formal modelling languages for interactive systems, including VDM \cite{vdm}, MAL \cite{MAL}, and PIM \cite{PIM}, as well as executable code (MISRA-C).
	
	\item {\bf The PVS back-end}. This includes the PVS theorem prover and the PVSio environment for model animation. The back-end is used for formal analysis of usability-related properties of the human-machine interface model, such as consistency of response to user actions and reversibility of user actions.
\end{itemize}

\subsection{Other tools}\label{sec:others}
MathWorks Simulink \cite{simulink} is a commercial tool for model-based design and analysis of dynamic systems. It provides a graphical model editor based on Statecharts, and functions for rapid generation of realistic prototypes. SCR \cite{heitmeyer1998scr} is a toolset for formal analysis of system requirements and specifications. Using SCR, it is possible to specify the behaviour of a system formally, use visual front-ends to demonstrate the system behaviour based on the specifications, and use a set of formal methods tools for the analysis of system properties. SCADE %\footnote{\url{http://www.esterel-technologies.com/}} 
and IBM's Rational Statemate %\footnote{\url{http://www-03.ibm.com/software/products/en/ratistat}} 
are two commercial tool sets for model-based development of interactive systems. The tool sets provide, among other features, rapid prototyping, co-simulation, and automated testing. Formal verification is supported by these tools, but is limited to the analysis of coding errors such as division-by-zero. Use-related requirements and tasks can be analysed only using simulation and testing. IVY \cite{ivy} is a workbench for formal modelling and verification of interactive systems. The tool provides developers with standard property templates that capture usability concerns in human-machine interfaces. The core of the IVY verification engine is the NuSMV model checker. A graphical environment isolates developers from the details of the underlying verification tool, thereby lowering the knowledge barriers for using the tool.
The particular tools that are of interest in the design and analysis of interactive systems enable the analysis of user activities, with a focus on what users do in terms of what they perceive about the systems and the actions they perform. Furthermore an important requirement for such tools is that the means of analysis and their results should be accessible to team members without a background in formal techniques, or even software development techniques.

\section{Case study and IDE showcase}\label{sec:example}\vspace*{-6pt}
The case study for comparing the selected tools is based on a subsystem of the Flight Control Unit (FCU) of the Airbus A380. It is an interactive hardware panel with several different buttons, knobs, and displays. The FCU has two main components: the Electronic Flight Information System Control Panel (EFIS CP), for configuring the piloting and navigation displays; and the Auto Flight System Control Panel (AFS CP), for setting the autopilot state and parameters. 

\begin{figure}[t]
	\centering
	\includegraphics[width=\textwidth]{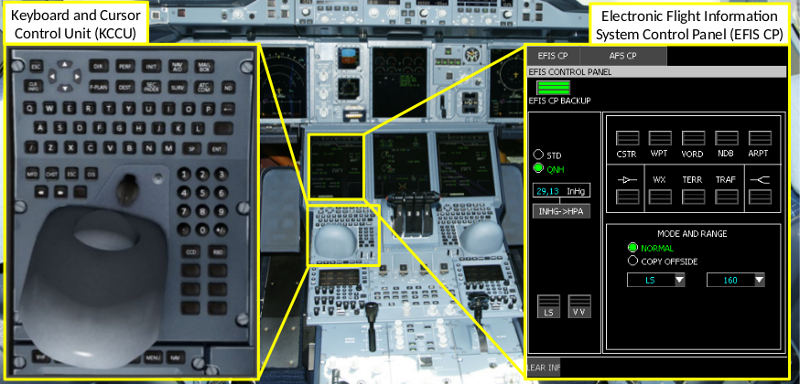}
	\caption{Keyboard and Cursor Control Unit and Flight Control Unit Software.}
	\label{fig:FCU}\vspace*{-8pt}
\end{figure}

In future cockpits, the interactive hardware elements of the FCU panel might be replaced by an interactive graphical application rendered on displays. This graphical software (hereafter, referred to as FCU Software) will provide the same functionalities as the corresponding hardware elements. This graphical software will be displayed on one of the screens in the cockpit. Pilots will interact with the FCU Software via the Keyboard and Cursor Control Unit (KCCU) that integrates keyboard and track-ball (see Figure~\ref{fig:FCU}).

The present paper illustrates how CIRCUS and PVSio-web can be used to create models and prototypes of the FCU Software. Developers can explore design options and analyse requirements for these future generation FCUs using these formal IDEs for model-based development of human-machine interfaces.
To keep the example simple, we focus further and analyse the EFIS CP. This component includes most of the fundamental interactive elements of the FCU.

\subsection{Description of the system and its use}\label{sec:desc}
A close up view of the EFIS CP is shown in the rightmost picture of Figure~\ref{fig:FCU}. The left panel of the EFIS CP window is dedicated to the configuration of the barometer settings (upper part) and of the Primary Flight Display (lower part). The right panel is dedicated to the configuration of the Navigation Display. The top part provides buttons for displaying navigation information on the cockpit displays. The ComboBoxes at the bottom allow the pilot to choose display modes and scale.

In this work, we focus more particularly on the configuration of the barometric settings (upper part of the left panel). This panel is composed of several widgets: two CheckButtons enable pilots to select either Standard (STD) or Regional Pressure Settings (QNH) mode. When in QNH mode, a number entry widget (EditBoxNumeric) enables pilots to set the barometric reference. Finally, a button (PushButton) enables pilots to switch the barometer units between inches of mercury (inHg) and hectopascal (hPa). When switching from one unit to the other, a unit conversion is triggered, and the barometer settings value on the display is updated accordingly. When the barometer unit is inHg, the valid range of values is [22, 32.48]. When the unit is hPa, the valid range is [745, 1100]. If the entered value exceeds the valid value range limits, the software automatically adjusts the value to the minimum (when over-shooting the minimum valid value) or the maximum (when overshooting the maximum valid value).

When starting the descent (before landing), pilots may be asked to configure the barometric pressure to the one reported by the airport. The barometric pressure is used by the altimeter as an atmospheric pressure reference in order to process correctly the plane altitude. To change the barometric pressure, pilots select QNH mode, then select the pressure unit (which depends on the airport), and then edit the pressure value in the EditBoxNumeric.

\subsection{Modelling and analysis using CIRCUS}
A prototype user interface was developed in CIRCUS that captures the functionalities of the FCU Software. The workflow for the modelling and analysis of this interactive system includes six main steps that are briefly detailed below.

\begin{figure}[t]
	\centering
	{\includegraphics[width=\linewidth]{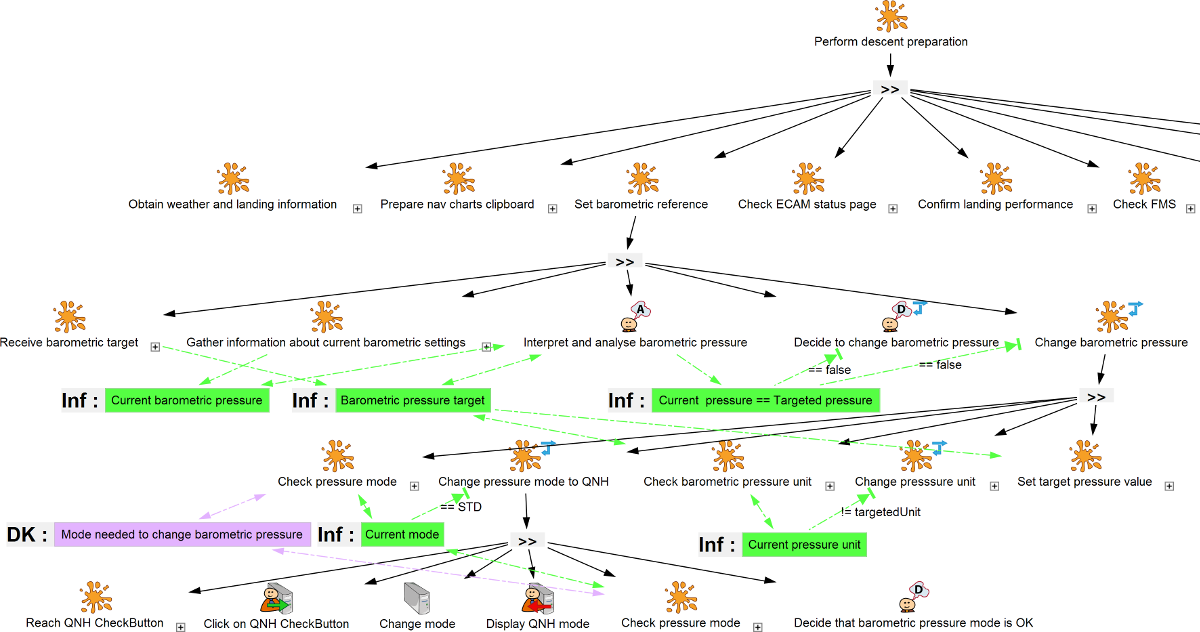}}
	\caption{Extract of the task model for the task ``Perform descent preparation''.}
	\label{fig:circus-task-model}\vspace*{-10pt}
\end{figure}

\smallskip\noindent
{\bf Step 1: Task analysis and modelling.~} This step describes user activities. It identifies goals, tasks, and activities that are intended to be performed by the operator. For example,  the task ``Perform descent preparation'', is decomposed into several HAMSTERS models represented as subroutines, components and instances of components~\cite{hcse2014}. Due to space constraints, we only describe an extract of the model in Figure~\ref{fig:circus-task-model}.

A  simplified version of the task model is described as the abstract task ``Perform descent preparation'' in the first row of Figure~\ref{fig:circus-task-model}. The second row refines this task into several abstract sub-tasks (e.g., ``Obtain weather and landing information''). Each one of these abstract tasks corresponds to a step of the operational procedure that is intended to be performed by the flight crew when preparing the descent phase~\cite{fcom}.

In the present paper, we focus on the ``Set barometric reference'' abstract task, refined in the third row. The task is decomposed as follows: the pilot receives the new barometric target (``Receive barometric target'' abstract task) and remembers the corresponding information (``Barometric pressure target''). The pilot then needs to gather information about the current barometer settings (``Gather information about current barometric settings''), thus remembering a new piece of information (``Current barometric pressure''). The pilot then needs to compare the two values that have been received in the previous two steps (``Interpret and analyse barometric pressure'' cognitive analysis task) creating the ``Current pressure == Targeted pressure'' information. If  the targeted pressure is different from its current value the pilot decides to change the pressure (``Decide to change barometric pressure'' cognitive decision task) and change it (``Change barometric pressure'' abstract task).

The fourth row of Figure~\ref{fig:circus-task-model} refines the ``Change barometric pressure'' abstract task as follows: the pilot must first check the pressure mode (``Check pressure mode'' abstract task), switch to QNH mode if the current mode is STD (``Change pressure mode to QNH'' abstract task), check the unit and change it if needed (``Check barometric pressure unit'' and ``Change pressure unit'' abstract tasks) and finally set the new pressure value (``Set target pressure value'' abstract task).

Finally, the fifth row refines the ``Change pressure mode to QNH'' abstract task. This task is performed by the pilot if the current mode is STD. In this task, the pilot first reaches the QNH checkButton (``Reach QNH CheckButton'' abstract task). Then s/he clicks on it (``Click on QNH CheckButton'' interactive input task). The system then changes the mode and displays the new state (``Change mode'' system task and ``Display QNH mode'' interactive output task). The pilot can then check that the new pressure mode is the good one (``Check pressure mode'' abstract task and ``Decide that barometric pressure mode is OK'' cognitive decision task). It is important to note that this task model is detailed both in terms of user task refinement (e.g., cognitive task analysis) to allow the analysis of workload and performance (see Step 3 below); and in terms of interactive task refinement (see, for instance, the refinement of the ``Change pressure mode to QNH'' abstract task which includes the ``Click on QNH Checkbutton'' interactive input task) to allow the compatibility assessment between the task model and the behavioural model of the system (see Step 6 below).

\smallskip\noindent
{\bf Step 2: Workload and performance analysis.~} As presented in Figure~\ref{fig:circus-task-model}, the HAMSTERS notation and tool enable human task refinement. It makes it possible to differentiate between cognitive, motor, and perception tasks as well as representing the knowledge and information needed by the user to perform a task. The refinement allows the qualitative analysis of user tasks, workload and performance. For example, the number of cognitive tasks and information that pilots need to remember may be effective indicators for assessing user workload~\cite{edcc2014}. This kind of analysis can be used to reason about automation design~\cite{smc2011}.

\smallskip\noindent
{\bf Step 3: User interface look and feel prototyping.~} This step aims at developing the user interface look and feel. A result of this step is described in the screen-shot of the EFIS CP presented in Figure~\ref{fig:FCU}. The widgets are organised in a style that is compatible with the library defined by the ARINC 661 standard~\cite{a661}.

\begin{figure}[t]
	\centering
	{\includegraphics[width=\linewidth]{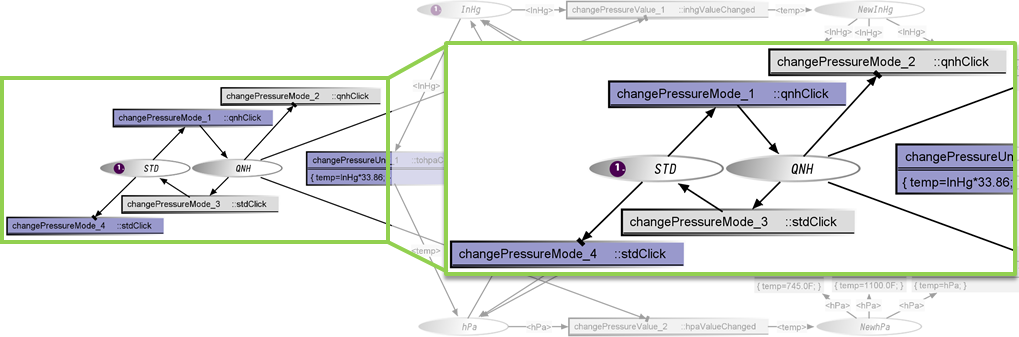}}
	\caption{ICO model of the barometer settings behaviour.}
	\label{fig:circus-ICO-model}\vspace*{-10pt}
\end{figure}

\smallskip\noindent
{\bf Step 4: User interface formal modelling.~} The behaviour of the user interface is specified using ICO models. The behaviour of the barometer settings part of the EFIS CP user interface is represented by the ICO model presented in Figure \ref{fig:circus-ICO-model}. The left part of this model (that has been enlarged) is dedicated to the pressure mode. As described in Section~\ref{sec:desc}, the pressure mode can be in two different mutually exclusive states: STD and QNH. The user can switch from one mode to the other one by clicking either STD or QNH CheckButton (clicking on a CheckButton while already in the corresponding mode is also possible but will have no impact on the pressure mode). This behaviour is defined by the enlarged part of the ICO model presented in Figure \ref{fig:circus-ICO-model}. The state of the pressure mode is represented by the presence of a token within ``QNH'' or ``STD'' places (in Figure \ref{fig:circus-ICO-model}, place ``STD'' holds a token meaning that the current pressure mode is STD). Transitions  ``changePressureMode\_1'' and ``changePressureMode\_2'' (resp. ``changePressureMode\_3'' and ``changePressureMode\_4'') correspond to the availability of the ``qnhClick'' (resp. ``stdClick'') event: when one of these two transitions is enabled, the ``qnhClick'' event is available (thus enabling the QNH CheckButton). The ``changePressureMode\_1'' transition therefore makes it possible to switch from STD pressure mode to QNH pressure mode as a result of clicking on the QNH CheckButton. The ``changePressureMode\_2'' transition allows the user to click on the QNH CheckButton while in QNH pressure mode without any impact on the pressure mode. The right part of the ICO model presented in Figure~\ref{fig:circus-ICO-model} (behind the enlarged part of the model) allows pressure to be changed. While we only present here a part of the ICO model describing the behaviour of the EFIS user interface, it is important to note that the ICO notation has also been used to describe the behaviour of the widgets and the window manager. The ICO models can be validated using the simulation feature.

\smallskip\noindent
{\bf Step 5: Formal analysis.~} The PetShop tool provides the means to analyse ICO models by the underlying Petri net model~\cite{fmis2013} using static analysis techniques as supported by the Petri net theory~\cite{peterson81}. The ICO approach is based on high level Petri nets. As a result the analysis approach builds on and extends these static analysis techniques. Analysis results must be carefully taken into account by the analyst as the underlying Petri net model can be quite different from the ICO model. Such analysis has been included in CIRCUS and can be interleaved with the editing and simulation of the model, thus helping to correct it in a style that is similar to that provided by  spell checkers in modern text editors~\cite{fomhci2016}). It is thus possible to check well-formedness properties of the ICO model, such as absence of deadlocks, as well as user interface properties, either internal properties (e.g., reinitiability) or external properties (e.g., availability of widgets). Note that it is not possible to express these user interface properties explicitly --- the analyst needs to express these properties as structural and behavioural Petri net properties that can be then analysed automatically in PetShop.

The analysis of the enlarged part of the ICO model presented in Figure~\ref{fig:circus-ICO-model} allows developers to check that, whatever action is taken, the pair of places ``STD'' and ``QNH'' will always hold one (and only one) token, exhibiting the mutual exclusion of the two states. Transitions connected to these places correspond to the availability of two events ``qnhClick'' and ``stdClick'', and therefore it can be demonstrated that these events will remain available whatever action is triggered. Lastly, there are two transitions in the model that correspond to the event ``qnhClick'' (transitions ``changePressureMode\_1'' and ``changePressureMode\_2''). This could potentially lead to non-determinism in the model. However, as ``changePressureMode\_1'' has place ``STD'' as input place and ``changePressureMode\_2'' has ``QNH'' place as input place, non-determinism is avoided due to the mutual exclusive marking of these places.

\smallskip\noindent
{\bf Step 6: Compatibility assessment between task models and user interface models.~} This step aims at guaranteeing that the task model and the formal model of the user interface behaviour are complete and consistent together (thus helping to guarantee that procedures followed by the operators are correctly supported by the system). A correspondence editing panel is used to establish the matching between interactive input tasks (from the task model) with system inputs (from the system model) and between system outputs (from the system model) with interactive output tasks (from the task model). The co-execution part of the SWAN tool provides support for validation as it makes it possible to find inconsistencies between the two models, e.g., sequences of user actions allowed by the system model and forbidden by the task model, or sequences of user actions that should be available but are not because of inadequate system design. The SWAN tool also provides support for automated scenario-based testing of an interactive application~\cite{eics2016}

%\smallskip\noindent
%{\bf Step 7: Mending of models or deployment.~} Finally, the last step consists in, following the results of the different analysis steps (workload and performance analysis, models simulation, formal analysis and compatibility assessment between task models and formal models) mending the different models to correct them or deploy the system.

%\input{content/pvsiowebAnalysis}
\subsection{Modelling and analysis using PVSio-web}
The focus of the PVSio-web analysis is the interaction logic of the EFIS data entry software. Here, we describe the modelling and analysis workflow supported by the tool, and highlight the main characteristics of the developed models (the full description is included as an example application in the PVSio-web tool distribution~\cite{cav2015}).

\begin{figure}[t]
	\centering
	{\includegraphics[width=\linewidth]{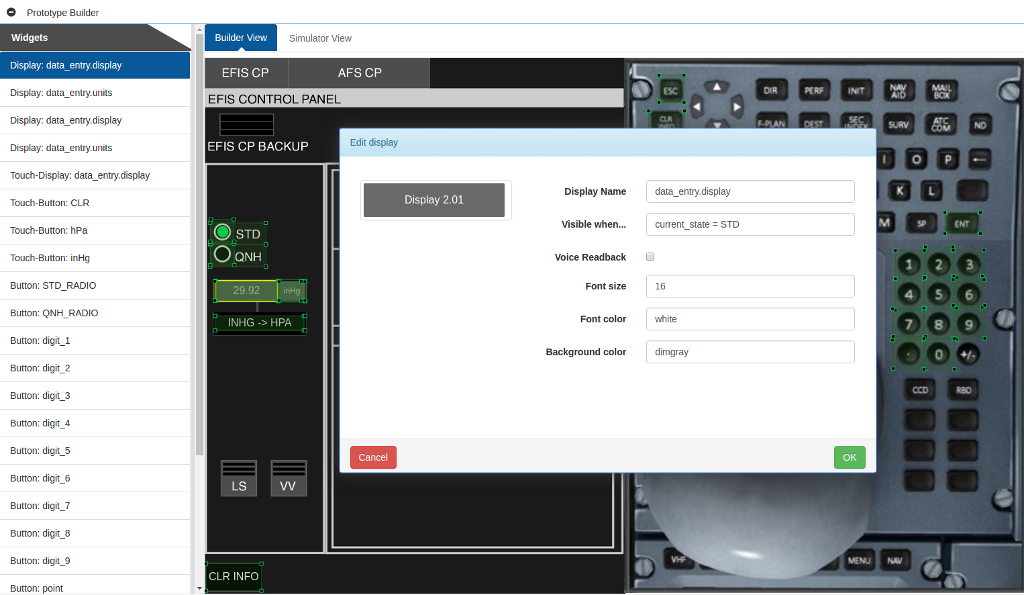}}
	\caption{FCU Software prototype developed in the PVSio-web Prototype Builder. Shaded areas over the picture identify interactive system elements.}
	\label{fig:EFIS-pvsioweb}\vspace*{-10pt}
\end{figure}

\begin{figure}[t]
	\centering
	\includegraphics[width=\linewidth,keepaspectratio]{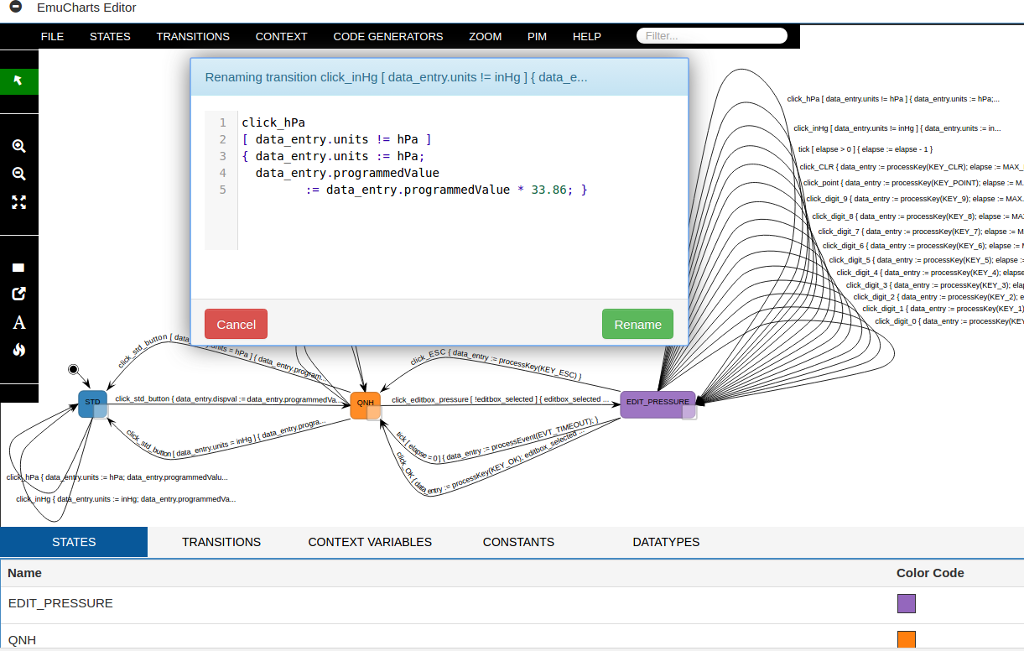}
	\caption{Emucharts model of the FCU Software created in the Emucharts Editor.}
	\label{fig:EFIS-emuchart}\vspace*{-10pt}
\end{figure}

\smallskip\noindent
{\bf Step 1: Define the visual appearance of the prototype.~} The visual aspect of the prototype is based on a picture of the EFIS panel. The PVSio-web Prototype Builder is used to create the visual appearance of the prototype and is defined using the Prototype Builder. A picture of the EFIS panel and KCCU are loaded in the tool, and interactive areas are created over relevant buttons and display elements (see Figure~\ref{fig:EFIS-pvsioweb}). Fifteen input areas were created over the picture of the Keyboard and Cursor Control Unit, to capture user actions over the number pad keys, as well as over other data entry keys  (ENT, CLR, ESC, and the units conversion button). Four display elements were created for rendering relevant status variables of the PVS model: a touchscreen display element handles user input on the EditBoxNumeric for entering the barometer pressure value; two display elements render the STD and QNH CheckButtons; an additional display element renders the pressure units.

\smallskip\noindent
{\bf Step 2: Define the behaviour of the prototype.~} The prototype is driven by a PVS model that includes an accurate description of the following features of the system: the modal behaviour of the data entry system; the numeric algorithm for units conversion; the logic for interactive data entry; and the data types used for computation (double, integer, Boolean). Modelling patterns were used to guide the development of the models (some of these patterns are described in \cite{harrison2016}). The model was developed using, in combination, the PVSio-web Emucharts Editor and the Emacs-based model editor of PVS. The Emucharts editor allowed us to create a statechart-based diagram that can be automatically translated into PVS models. The Emacs-based model editor was used to build a library function linked to the Emucharts diagram to improve modelling efficiency. The developed Emucharts (shown in Figure \ref{fig:EFIS-emuchart}) includes the following elements: 3 states (STD, QNH, and EDIT PRESSURE) representing three different modes of operation; 25 transitions, representing the effect of user actions on the Keyboard and Cursor Control Unit when adjusting the barometer settings, and internal timers handling timeouts due to inactivity during interactive data entry; and 9 status variables, representing the state of the system (units, display value, programmed value, etc.). The library function, ProcessKey, is used within the Emucharts diagrams to define the effect on state variables of transitions associated with key presses.

\smallskip\noindent
{\bf Step 3: Model validation.~} This analysis ensures internal consistency of the model, as well as checking accuracy with respect to the real system. Internal consistency is assessed by discharging proof obligations (called type-check-conditions) automatically generated by the PVS theorem prover. These proof obligations check coverage of conditions, disjointness of conditions, and correct use of data types. For the developed Emucharts model, PVS generated 22 proof obligations, all of which were discharged automatically by the PVS theorem prover. % (see Figure \ref{fig:tccs}).
Accuracy of the model is assessed by using the prototype to engage with Airbus cockpit experts. Experts can press buttons of the prototype, and watch the effect of interactions on the prototype displays. By this means it is possible to check that the prototype behaviour resembles that of the real system.

\smallskip\noindent
{\bf Step 4: Formal analysis.~} The prototype and the PVS theorem prover are used in combination to analyse the model. The prototype is used to perform a lightweight formal analysis suitable to establish a common understanding within a multidisciplinary team of the correct interpretation of safety requirements and usability properties. This analysis consists in the execution of sample input key sequences demonstrating scenarios where a given requirement is either satisfied or fails. This initial lightweight analysis based on test cases is extended to a full formal analysis using the PVS theorem prover, to check that requirements and properties of the model are satisfied for all input key sequences in all reachable model states. To perform this full analysis, PVS theorems need to be defined that capture the requirements and properties. They are expressed using structural induction and a set of templates described in \cite{HarrisonMC:16}. An example property that can be analysed is {\em consistency of device response to user actions}. The consistency property is motivated by the fact that users quickly develop a mental model that embodies their expectations of how to interact with a user interface. Because of this, the overall structure of a user interface should be consistent in its layout, screen structure, navigation, terminology, and control elements. Example consistency properties are: a designated set of function buttons always change the mode; a further set of keys, for example concerned with number entry, will always change the barometric variable relevant to the mode but do not change the mode; an \Mword{enter} key always changes the relevant parameter when in the relevant mode; an \Mword{escape} key ensures that the value set in the mode is discarded and the barometric value  reverts to the value it had when it entered the mode.

In PVS, the consistency template is formulated as a property of a group of actions $A_c \subseteq \wp(S \rightarrow S)$, or it may be the same action under different modes, requiring that all actions in the group have similar effects on specific state variables selected using a filter. The relation $\mathit{consistent}: C \cross C \rightarrow T$  connects a filtered state, before an action occurs (captured by $\mathit{filter\_pre}: S \cross MS \rightarrow C$), with a filtered state after the action (captured by $\mathit{filter\_post}: S \cross MS \rightarrow C$). The description of the filters and the \Mword{consistent} relation specify the consistency across states and across actions. Here $MS$ is defined to be a set of modes. Two modes are relevant here. A set of modes not defined includes the mode that allows the entry of the barometer value. Within the barometer entry mode are two modes that relate to the different units that can be used to enter the barometric values, defined as: inHg and hPa.
A general notion of consistency assumes that the property is restricted to a set of states focused by a guard:  $\mathit{guard}: S \cross MS \rightarrow T$. This guard may itself be limited by a mode. The general consistency template can therefore be expressed as:

\smallskip
%\begin{mdframedbox}{Consistency}	
\begin{tcolorbox}{\bf Consistency}
	\begin{Mbit}
		\Mplain{\forall a \in A_c \subseteq \wp(S \rightarrow S) , s \in S, m \in MS:}\\
		\>\Mplain{\mathit{guard}(s, m) \land \mathit{consistent}(\mathit{filter\_pre}(s, m),\mathit{filter\_post}(a(s), m))}
	\end{Mbit}
\end{tcolorbox}
%\end{mdframedbox}\smallskip

%\noindent
%A consistency property may require that in a given mode (defined in the \Mword{guard}), specific state variables (defined in the filters) are never changed, or alternatively always changed.

\noindent
Two examples are now used to illustrate the use of the consistency template. The first is that a set of actions never change the barometric entry mode. The \Mword{pre\_filter} and \Mword{post\_filter} both extract the barometric entry mode, and are of the form
\texttt{filter\_baro(st:~state):~UnitsType = Units(st)}. This property relates directly to modes and therefore the mode parameter can be omitted in the filter definition. The set of actions determined by {\tt state\_transitions\_actions} which encompasses the set of transitions as determined by the enabled actions in the barometric mode. In summary, the \Mword{consistent} relation in this case is equality and the theorem that instantiates the consistency template is:

{
	\begin{verbatim}
	modeinvariant: THEOREM FORALL (pre, post: state): 
	state_transitions_actions(pre, post) => (filter_baro(pre) = filter_baro(post))
	\end{verbatim}}

\noindent
On the other hand the action \Mword{click_{hPa}} always changes the entry mode. So here \Mword{consistent} is inequality.

{\begin{verbatim}
alwayschgmode: THEOREM
  FORALL (pre, post: state):
    (post = click_hPa(pre) AND guard_baro(pre))
         => filter_baro(pre) /= filter_baro(post)
\end{verbatim}}

\section{Tool comparison}\label{sec:comparison}
In this section, we first present the criteria that were identified to compare the characteristics and functionalities of the two tools. These criteria form a basis for the comparison of these two tools.

\subsection{Comparison criteria}
We identified 22 criteria suitable to compare the characteristics and functionalities of the two tools. These criteria are general, and can be used as a reference to define a taxonomy suitable to classify and compare other similar formal IDEs for user interface modelling and analysis. These criteria are divided in four categories:

\begin{itemize}
	\item {\bf General aspects of the tools}
	\begin{enumerate}
		\item[1.] Scope/purpose of the tool within the development process, e.g., requirements analysis, prototyping, verification.
		\item[2.] Tool features, e.g., modelling of user tasks and goals, analysis of usability properties, simulation of user tasks.
		\item[3.] Tool extensibility, e.g., to model systems from different application domains, or to perform a different type of analysis
		\item[4.] Prerequisites and background knowledge, e.g., distributed systems, object oriented languages, Petri Nets, task modelling, PVS.
		\item[5.] IDE instance and principle, e.g., Eclipse plugin, Web, Netbeans API
		\item[6.] IDE availability, e.g., snapshot, demo, downloadable, open source.
	\end{enumerate}
	
	\item {\bf Modelling features}
	\begin{enumerate}%[resume]
		\item[7.] Notation names, e.g., ICO, HAMSTERS, Emucharts, PVS.
		\item[8.] Notation instance, e.g., Petri Net, state machines, higher-order logic.
		\item[9.] Notation paradigm, e.g., event-based, state-based, declarative.
		\item[10.] Structuring models, e.g., object-oriented, functional, component-based.
		\item[11.] Model editing features, e.g., textual, visual, autocompletion support.
		\item[12.] Suggestions for model improvements, e.g., strengthening of pre-conditions.
	\end{enumerate}
	
	\item {\bf Prototyping features}
	\begin{enumerate}%[resume]
		\item[13.] Support for prototype building, e.g., visual editor, library of widgets.
		\item[14.] Execution environment of the prototype, e.g., Java virtual machine, Javascript execution environment.
		\item[15.] User interface testing, e.g., automatic generation of input test cases.
		\item[16.] Human-machine interaction techniques, e.g., Pre-WIMP (input dissociated from output), WIMP, post-WIMP, tangible, multimodal.
		\item[17.] Code generation, e.g., C, C++, Java.
	\end{enumerate}
	
	\item {\bf Analysis of human-machine interaction}
	\begin{enumerate}%[resume]
		\item[18.] Verification type, e.g., functional verification, performance analysis, hierarchical task analysis;
		\item[19.] Verification technology, e.g., theorem proving, static analysis.
		\item[20.] Scalability of the analysis, e.g., illustrative examples, industrial size.
		\item[21.] Support for the analysis of the wider socio-technical system.
		\item[22.] Related development process, e.g., user centered design, waterfall development process, agile development.
	\end{enumerate}
\end{itemize}

\subsection{CIRCUS and PVSio-web comparison}
In this section, we discuss, following the four categories of criteria identified above, the comparison of CIRCUS and PVSio-web. A detailed assessment of all the criteria presented above is presented in tabular form in the Appendix.

\smallskip\noindent
{\bf General aspects of the tools.~}
From a high-level perspective, the scope of CIRCUS and PVSio-web is the formal development of user interfaces. Both tools support modelling and analysis of the interaction logic of the user interface software. However, the two tools offer different modelling and analysis technologies that are tailored to support two different (and complementary) styles of assessment of user interfaces. CIRCUS supports explicit modelling of user tasks and goals, allowing developers to simulate user tasks and check their compatibility with the interactive behaviour of the system. PVSio-web, on the other hand, supports explicit modelling of general usability and safety properties, facilitating the assessment of compliance of a user interface design with design guidelines and best design practices (e.g., according to standards or regulatory frameworks). Whilst a certain level of background knowledge is needed to use the tools effectively, basic knowledge about Petri nets and task models (for CIRCUS) and state machines and state charts (for PVSio-web) is already sufficient to get started with illustrative examples. This is extremely useful to reduce the typical knowledge barriers faced by novice users. The two IDEs are developed using standard technologies supported by multiple platforms (Netbeans Visual API for CIRCUS, Web technologies for PVSio-web), and can be executed on any standard desktop/laptop computer.

\smallskip\noindent
{\bf Modelling features.~}
Both tools provide powerful graphical IDEs designed to assist developers in the creation of formal models. CIRCUS uses specialised graphical notations and diagrams: the ICO notation is used for building system models; the HAMSTERS notation is used for describing user tasks. ICOs are based on object-oriented extensions to Petri nets, and support both event-based and state-based modelling. HAMSTERS is a procedural notation. The complexity of models is handled using information hiding (as in object-oriented programming languages), and component-based model structures. This facilitates the creation of complex models, as well as the implementation of editing features that are important for developers, such as auto-completion of models and support for parametric models. The use of specific notations, however, limits the ability of developers to import external models created with other tools, or export CIRCUS models to other tools. PVSio-web, on the other hand, uses modelling patterns to support the modelling process. Developers can use either a graphical notation (Emucharts diagrams, or a textual notation (PVS higher-order logic), or a combination of both, to specify the system model. This has many benefits: software developers that are familiar with Statecharts can build models using a language that is familiar for them, and gradually learn PVS modelling by examples, checking how the Emucharts model translates into PVS; Emucharts models can be translated into popular formal modelling languages different than PVS (e.g., VDM); expert PVS users can still develop entire models using PVS higher-order logic only, and software developers can import these PVS models as libraries, thus facilitating model re-use. The main drawback is that the current implementation of Emucharts lacks mechanisms for model optimisation (e.g., a battery of similar PVS functions is generated instead of a single function with a parameter), and technical skills are necessary to understand model improvements suggested by the tool (through the PVS type-checker).

\smallskip\noindent
{\bf Prototyping features.~}
Both IDEs provide a visual editor for rapid generation of prototypes supporting a range of interaction styles, including: graphical user interfaces with windows, icons, menus, and pointer (WIMP); user interfaces with physical buttons (pre-WIMP); touchscreen-based user interfaces (post-WIMP); and multi-modal user interfaces (e.g., providing both visual and auditory feedback). Both tools promote the use of the Model-View-Controller (MVC~\cite{krasner1988description}) paradigm, with a clear separation between the visual appearance of the prototype and the logic behaviour. Whilst prototypes developed with the two IDEs share these similarities, prototype building and implementation is substantially different in the two IDEs. CIRCUS prototypes are developed in Java (for their visual appearance) and in ICO models (for their behaviour). Developers can define their own widgets library. For example, for the case study presented in Section~\ref{sec:example}, we created a library of widgets whose visual aspect and behaviour is compatible with that described in the ARINC 661 standard. PVSio-web prototypes are developed in JavaScript, and their behaviour is defined by a PVS executable model. Rapid prototyping is enabled by a lightweight building process where the visual aspect of the prototype is defined by a picture of the real device, virtually reducing to zero the time and effort necessary to define the visual appearance of the prototype. Initial support for code generation is also available for MISRA-C, for behavioural models developed using Emucharts~\cite{mauro2016}. A specialised tool (Prototype Builder) is provided with the IDE, to facilitate the identification of interactive areas over the picture, and to link these areas to the PVS model. The current implementation of the Prototype Builder supports only the definition of push buttons and digital display elements, and developers need to edit a JavaScript template manually to introduce more sophisticated widgets (e.g., knobs, graphical displays, etc.). Integration of these more sophisticated widgets in the Prototype Builder is currently under development.

\smallskip\noindent
{\bf Analysis of human-machine interaction.~}
Multiple verification technologies are used in the two IDEs to enable the efficient analysis of human-machine interaction. Both tools build on established formal methods technologies, and enable lightweight formal analysis based on simulation and testing. CIRCUS implements static analysis techniques from Petri nets theory to perform automatic analysis of well-formedness properties of the model (absence of deadlocks, token conservation), and of basic aspects of the interactive system design (e.g., reinitiability of the user interface and availability of widgets). Simulation is used for functional analysis and quantitative assessment of the system. Either direct interaction with the prototype and automated execution of task models can be used during simulations. Properties verified by this means include: compliance with task models; statistics about the total number of user tasks, and estimation of the cognitive workload of the user based on the types of human-machine interactions necessary to operate the system. PVSio-web uses the standard PVS theorem proving system to analyse well-formedness properties of the model (coverage of conditions, disjointness of conditions, and correct use of data types). Usability and safety requirements can be verified using both lightweight formal verification and full formal verification. Lightweight verification is based on interactive simulations with the prototypes. User interactions can be recorded and used later as a basis for automated testing in a way similar to the way task models are used in CIRCUS. Full formal verification is carried out in the PVS theorem prover, and is partially supported by property templates capturing common usability and safety requirements described in the ANSI/AAMI/IEC HF75 usability standard. Although the full formal analysis is in general not fully automatic, the combined use of property templates and modelling patterns usually leads to proof attempts where minimal human intervention is necessary to guide the theorem prover (typically, for case-splitting and instantiation of symbolic identifiers). Proof tactics for full automatic verification of a standard battery of property templates are currently under development. Dedicated front-ends presenting verification results in a form accessible to human factors specialists are also being investigated.

\section{Conclusion and perspectives}\label{sec:conclu}
In this paper, we presented a first step towards providing guidance to developers to understand which formal tool can be used most effectively for which kind of analysis of interactive systems. This is achieved through the identification of 22 criteria enabling the characterisation of IDEs for interactive systems formal prototyping and development. These criteria have been used to compare two state-of-the-art formal tools developed by two different research teams: CIRCUS, a toolkit for model-based development of interactive systems; and PVSio-web, a toolkit for model-based development of user interface software based on the PVS theorem proving system. In order to assess all the criteria, we modelled and analysed a case study from the avionics domain using these two tools. The result of this comparison led to the conclusion that the two studied tools are complementary rather than competitive tools. Whilst they have roughly the same scope (formal development of user interfaces), these two tools enable different kinds of modelling and analysis. For instance, CIRCUS supports explicit modelling of user tasks and goals, allowing developers to simulate user tasks and check their compatibility with the interactive behaviour of the system while PVSio-web supports explicit modelling of general usability and safety properties, facilitating the assessment of compliance of a user interface design with design guidelines and best design practices. These two analysis styles are complementary, and both provide important insights about how to develop high-confidence user interfaces. Based on this understanding, we are now developing means to integrate the two IDEs, to enable new powerful analysis features, such as automated scenario-based testing of user interfaces~\cite{eics2016}. The envisioned integration is introduced at two levels: at the modelling level, developing PVSio-web extensions for importing/translating HAMSTERS task models into PVS models and properties; and at the simulation level, building CIRCUS extensions for co-execution of task models and PVSio-web prototypes. Additional extensions under development for the two toolkits include: modelling patterns for describing human-machine function allocation; proof tactics and complementary use of different verification technologies for improved automation of usability and safety properties; innovative front-ends for inspecting formal proofs supporting safety and usability claims of user interfaces; and widgets libraries for different application domains.

{\footnotesize
~\\\noindent
{\bf Acknowledgment.~}
This work is partially supported by: Project NORTE-01-0145-FEDER-000016, financed by the North Portugal Regional Operational Programme (NORTE 2020), under the PORTUGAL 2020 Partnership Agreement, and through the European Regional Development Fund (ERDF); Conselho Nacional de Desenvolvimento Cient\'ifico e Tecnol\'ogico (CNPq) PhD scholarship.
}

\bibliographystyle{eptcs}\bibliography{F-IDE2016}

\newcounter{comp} \setcounter{comp}{0}
\newcommand{\smallfont}{\fontsize{8}{8}\selectfont}
\newcommand{\comparison}[3]{\stepcounter{comp}
	{\smallfont\thecomp.} & {\smallfont #1}& {\smallfont #2}& {\smallfont #3}\\\hline
}
\newcommand{\continuesOnNextPage}{
	& & {\smallfont\em continues on next page...} & \\\hline
}
\newcommand{\heading}{
	& {\smallfont\bf Formal IDE}& {\smallfont\bf CIRCUS}& {\smallfont\bf PVSio-web}\\\hline
}
\setlength{\tabcolsep}{4pt}

\begin{landscape}
	\begin{table}[]
		{\bf Appendix} \\~\\
		\setlength{\aboverulesep}{0pt}\setlength{\belowrulesep}{0pt}\setlength{\extrarowheight}{0.75ex}
		\begin{tabular}{@{}c>{\raggedright}p{0.14\linewidth}p{0.36\linewidth}p{0.36\linewidth}@{}}\toprule[1.5pt]
			\heading
			
			\comparison{Scope/purpose}{Interactive system prototyping, development, and analysis.}{User interface software prototyping and analysis.}
			
			\comparison{Tool features}{User task and goals description, interaction logic (dialog) and interaction techniques modelling, interactive system prototyping, support for verification of properties, assessment of compatibility between user tasks and interactive system prototype.}{Interaction logic modelling, rapid prototyping of user interface software, verification of safety requirements and usability properties, code generation and documentation.}
			
			\comparison{Tool extensibility}{Each tool within CIRCUS offers an API supporting connection to other computing systems. For instance, connecting PetShop execution engine to cockpit software simulators or connecting Petri net analysis tools to PetShop analysis module.}{PVSio-web has a plug-in based architecture that enables the rapid introduction of new modelling, prototyping, and analysis tools; support for new widgets types and widgets libraries can be introduced in Prototype Builder; Emucharts Editor can be extended with new model generators and code generators.}
			
			\comparison{Background knowledge}{Object-Oriented Petri Nets (for Petshop), Java programming, distributed systems, hierarchical task modelling.}{State machines, PVS higher order logic and PVS theorem proving (only required for full formal verification).}	
			
			\comparison{IDE principles}{Netbeans Visual API}{Web}
			
			\comparison{IDE availability}{Available upon request for collaborations only.}{Open source, downloadable at \url{http://www.pvsioweb.org}}
			
			\comparison{Notation names}{ICO, HAMSTERS.}{Emucharts, PVS.}
			
			\comparison{Notation instance}{Petri Net, task models.}{Statecharts, higher-order logic.}
			
			\comparison{Notation paradigm}{Event-based, state-based, procedural.}{Event-based, state-based, functional.}
			
			\comparison{Structuring models}{Object-oriented, component-based.}{Module-based.}
			
			\comparison{Model editing features}{Graphical editing of task models, ICO models and their correspondences, auto-completion features of models, visual representation of properties on models, simulation of models at editing time.}{Graphical and textual editing of models, automatic generation of PVS models.}
			
			\comparison{Suggestions for model improvement}{Suggestions for model correction by real time analysis of models and continuous visualization of analysis results.}{Strengthening of pre- and post- conditions of transition functions (based on proof obligations generated by PVS).}
			
			\continuesOnNextPage
			\bottomrule[1.5pt]
		\end{tabular}\end{table}\end{landscape}
		
		\begin{landscape}
			\begin{table}[]
				\setlength{\aboverulesep}{0pt}\setlength{\belowrulesep}{0pt}\setlength{\extrarowheight}{0.75ex}
				\begin{tabular}{@{}c>{\raggedright}p{0.14\linewidth}p{0.36\linewidth}p{0.36\linewidth}@{}}\toprule[1.5pt]
					
					\heading
					\comparison{Prototype building}{Use of graphical user interface editor of NetBeans for standard interactions (e.g. WIMP), possible to create interactive components and assemble them for non standard intereactions (e.g. multitouch).}{Visual editing, based on a picture of the real system.}
					
					\comparison{Prototype execution}{Java Virtual Machine.}{Javascript execution environment, Lisp.}
					
					\comparison{User interface testing}{Automatic execution of test sequences based on a task model}{Automated execution of input test sequences recorded during interactions with the prototype.}
					
					\comparison{Human-machine interaction techniques}{Pre-WIMP, WIMP,  post-WIMP, multimodal, multi-touch. Run-time re-configuration of interaction techniques.}{Pre-WIMP, WIMP, post-WIMP, multimodal.}
					
					\comparison{Code generation}{Run-time execution of ICO models (to support prototyping and co-execution of task and system models).}{Run-time execution of PVS executable models through the PVS ground evaluator (to support rapid prototyping), and automatic generation of production code compliant to MISRA-C (only for formal models developed using Emucharts diagrams).}
					
					\comparison{Verification types}{Well-formedness of the model: absence of deadlocks, token conservation.
						Functional analysis: reinitiability; availability of widgets; compliance with task models. Quantitative analysis: statistics about the total number of user tasks; estimation of the cognitive workload of the user based on the types of human-machine interactions necessary to operate the system. Simulation-based analysis through model animation.}{Functional analysis, including: coverage of conditions, disjointness of conditions, correct use of data types, compliance with design requirements. Simulation-based analysis through model animation.}
					
					\comparison{Technology}{Static analysis of Petri Nets; interactive simulation of task and system models. Proofs and properties verification left to the analyst.}{Theorem proving; interactive simulations.}
					
					\comparison{Scalability}{Applied to very large scale (industrial) applications (more than 200 models).}{User interface prototype of stand-alone devices.}
					
					%\continuesOnNextPage
					%\bottomrule[1.5pt]
					%\end{tabular}\end{table}\end{landscape}
					%
					%\begin{landscape}
					%	\begin{table}[]
					%		\setlength{\aboverulesep}{0pt}\setlength{\belowrulesep}{0pt}\setlength{\extrarowheight}{0.75ex}
					%		\begin{tabular}{@{}c>{\raggedright}p{0.14\linewidth}p{0.36\linewidth}p{0.36\linewidth}@{}}\toprule[1.5pt]
					%			
					%\heading
					
					\comparison{Analysis of the wider socio-technical system}{Modelling of integrated views of the three elements of socio-technical systems (organization, human and interactive systems); however, FRAM-based description of organization and variability of performance has only be addressed at model level and not a tool level.}{Modelling patterns based on distributed cognition theory have been explored in PVS but are not currently integrated in the IDE.}
					
					\comparison{Related development process}{User centered design (task-based design), iterative development, model-based engineering.}{User centered design, agile development, model-based engineering.}
					\bottomrule[1.5pt]
				\end{tabular}\end{table}\end{landscape}

\end{document}